\newif\ifpdf
   \ifx\pdfoutput\undefined
   \pdffalse  
\else
   \pdfoutput=1  
   \pdftrue
\fi

\documentclass[11pt,a4paper]{article}
\usepackage[super,comma,sort&compress]{natbib}

\ifpdf
\usepackage[pdftex]{graphicx} 
\else
\usepackage[dvips]{graphics} 
\fi

\begin{document}
\begin{center}
{\bf\large Electrons and Phonons on the Square Fibonacci Tiling}

Roni Ilan, Edo Liberty, Shahar Even-Dar Mandel, and Ron Lifshitz.\\
School of Physics and Astronomy, Raymond and Beverly Sackler
  Faculty\\ of Exact Sciences, Tel Aviv University, Tel Aviv 69978, Israel.
\end{center}

\abstract{We study the Schr\"{o}dinger equation for an off-diagonal
  tight-binding hamiltonian, as well as the equations of motion for
  out-of-plane vibrations on the separable square Fibonacci
  quasicrystal. We discuss the nature of the spectra and wave
  functions of the solutions. }

\section{\normalsize  Separable Quasicrystal Models}

Much research has been conducted on the properties of excitations in
quasicrystals, ever since the early interest in this
question\cite{kohmoto} and up to this day.\cite{reviews} Many exact
results have been found for 1-dimensional ($1d$) quasicrystals, yet
the properties of excitations in $2d$ and $3d$ quasicrystals are known
to a much lesser degree. Furthermore, the different models that have
been mostly studied---such as $1d$ chains, the standard decagonal,
octagonal, and icosahedral tilings, or actual structural models of
real quasicrystals---do not allow one to easily focus on the
dimensional dependence of the physical properties of quasicrystals,
though some interesting heuristic arguments have been given by
Sire.\cite{sire}

One of us~\cite{squarefib} has recently studied the geometric
properties and calculated the diffraction pattern of the square and
cubic Fibonacci tilings, suggesting that they be used as models for
$2d$ and $3d$ quasicrystals (with obvious generalization to any higher
dimension).  The advantage of these prototypical models of
$d$-dimensional quasicrystals is that they are separable---certain
problems, such as the eigenvalue problems studied here, can be
decomposed into $d$ separate one-dimensional problems, yielding
straightforward solutions, while allowing directly to focus on the
effect of dimensionality on the problem being studied. The most
apparent disadvantage of these models is that they do not occur in
``real'' quasicrystals, yet they should not be dismissed as irrelevant
because they can be artificially constructed using, for example,
conducting nanowires, coupled nanomechanical resonators, or photonic
quasicrystals.

\begin{figure}[tb]
\hspace*{0pt}
\begin{center}

\scalebox{0.20}{\rotatebox{00}{\includegraphics*{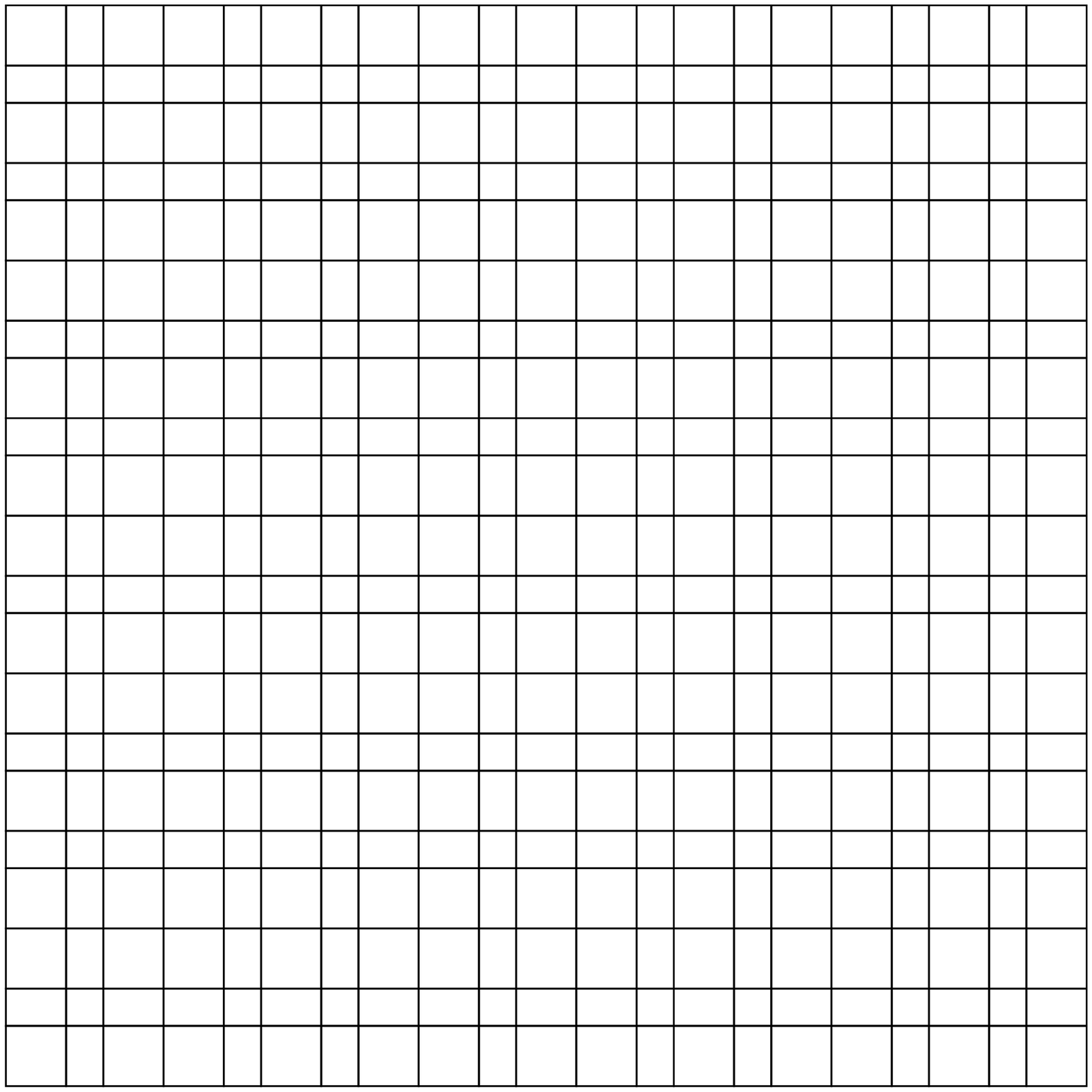}}}\\
\parbox{10.cm}{\caption{A section of the square Fibonacci tiling.}}

\end{center}
\end{figure} 

The square Fibonacci tiling is constructed by taking two identical
grids---each consisting of an infinite set of lines whose inter-line
spacings follow the well-known Fibonacci sequence of short ($S$) and
long ($L$) distances---and superimposing them at a $90^{\circ}$ angle,
as shown in Fig.~1. This construction can be generalized, of course,
to any quasiperiodic sequence as well as to higher dimensions. If the
original $1d$ sequence has inversion symmetry the generated $2d$ and
$3d$ quasicrystals will have square and cubic point group symmetry,
respectively.

\section{\normalsize  Electrons and (Scalar) Phonons}

Here we take advantage of the separability of the square (and cubic)
Fibonacci tilings to study two problems: (1) The tight binding
hamiltonian with zero on-site energy and hopping amplitudes $t$ for
vertices connected by long ($L$) edges, $1$ for vertices connected by
short ($S$) edges, and zero for vertices that are not connected by
edges; and (2) The normal modes of out-of-plane vibrations of a
network of unit masses connected by springs according to the square
Fibonacci tiling, with spring constants $k$ for long edges, and $1$
for short edges.

The 2-dimensional Schr\"odinger equation for the connected-neighbor
tight-binding hamiltonian is given by
\begin{eqnarray}
  \label{eq:twoDeq}\nonumber
  &t_{n+1} \Psi(n+1,m) + t_n \Psi(n-1,m)\hskip3cm\\
  &+ t_{m+1} \Psi(n,m+1) + t_m \Psi(n,m-1)
  = E\Psi(n,m),
\end{eqnarray}
where $\Psi(n,m)$ is the value of a $2d$ eigenfunction on a
vertex labeled by the two integers $n$ and $m$, and $E$ is the
corresponding eigenvalue. The hopping amplitudes $t_j$ are equal to 1
or $t$ according to the Fibonacci tiling as described above. The
equations of motion for the network of springs is obtained from the
Schr\"odinger equation~(\ref{eq:twoDeq}) by replacing $t_j$ by the
spring constant $k_j$, replacing $E$ by $(k_{n+1}+k_n+k_{m+1}+k_m) -
\omega^2$, where $\omega$ is the normal frequency, and viewing
$\Psi(n,m)$ as the out-of-plane displacement of the $(n,m)$ vertex.
The generalization of Eq.~(\ref{eq:twoDeq}) to three or any higher
dimension is obvious.

With no additional assumptions other than the absence of diagonal
hopping or diagonal springs this 2-dimensional eigenvalue problem, as
well as its higher dimensional versions, are all separable.
Two-dimensional eigenfunctions can be expressed as products
\begin{equation}
  \label{eq:product}
  \Psi_{ij}(n,m)=\psi_i(n)\psi_j(m),
\end{equation}
with eigenvalues
\begin{equation}
  \label{eq:sum}
  E_{ij}=E_i + E_j, \qquad \textrm{or} \qquad 
   \omega_{ij}^2=\omega_i^2+\omega_j^2,
\end{equation}
where $\psi_i(n)$ and $\psi_j(n)$ are two of the eigenfunctions of the
corresponding $1d$ eigenvalue equation on the Fibonacci
chain, with eigenvalues $E_i$ and $E_j$ or $\omega_i^2$ and
$\omega_j^2$, respectively.

Separable quasiperiodic hamiltonians have been studied on such models
in the past,~\cite{ueda} mainly focusing on square or cubic periodic
lattices with constant hopping amplitudes (or springs) and on-site
energies (or masses) that follow the same quasiperiodic sequence in
all directions. These so-called ``diagonal'' models are separable if
one requires the on-site energies (or masses) to be of the form
$V(n,m,\ldots) = V_x(n) + V_y(m) +\ldots$, in which case they have
solutions as above.

\section{\normalsize Adding Energies}

When solving the $1d$ tight-binding problem using the standard methods
of transfer matrices and trace maps,~\cite{kohmoto} one finds for any
$t$ that the $n^{th}$ approximant, with $F_n$ atoms per unit cell, has
a spectrum containing $F_n$ continuous bands, where $F_n$ is the
$n^{th}$ Fibonacci number. As $n$ increases and the quasicrystal is
approached, the bands become narrower, and in the limit $n\to\infty$
the spectrum becomes singular-continuous, containing a Cantor-like set
of points whose total bandwidth (Lebesgue measure) is zero.
Figs.~2(a)--(c) show these bands for the first few approximants for
three values of the parameter $t$. One sees that as $t$ increases away
from the periodic ($t=1$) structure the gaps that are formed become
increasingly wider.

When adding two such spectra as in Eq.~(\ref{eq:sum}) we first note
that due to the degeneracy of $E_{ij}=E_{ji}$ the maximum number of
bands for the $n^{th}$ approximant, if there is no overlap at all, is
$F_n(F_n+1)/2$. The degree of overlap depends on $t$. We find three
distinct behaviors which can be described qualitatively as follows:
(1) For large enough $t$ the gaps in the $1d$ spectrum are
sufficiently wide such that the $2d$ spectrum contains well separated
bands [Fig.~2(f)]. As $n$ increases the total number of bands
increases and in the limit $n\to\infty$ the $2d$ spectrum becomes
singular-continuous with zero total bandwidth. (2) For $t$ close to 1,
gaps in the $1d$ spectrum are sufficiently small such that the $2d$
bands greatly overlap, forming a single or a few bands even in the
$n\to\infty$ limit [Fig.~2(d)]. The spectrum of the $2d$ quasicrystal
is therefore absolutely continuous.  (3) For intermediate $t$ a
peculiar situation exists (observed also in the diagonal
models~\cite{ueda}) where even though the total number of bands
increases with $n$ the total integrated bandwidth tends to a finite
value [Fig.~2(e)]. It is not clear to us at this point whether the
spectrum of the $2d$ quasicrystal in this case contains an absolutely
continuous part or whether it remains singular-continuous as in $1d$.

Clearly, the transitions between the different behaviors are pushed
to higher values of $t$ as the dimension increases and additional $1d$
spectra are added. We shall provide elsewhere a detailed analysis of
the phase diagram of this system as a function of $t$ and dimension.
We shall also describe the differences between electrons and
phonons.

\section{\normalsize Multiplying Wave Functions}

An intriguing paradox arises in light of the discussion above. Since
for any $t$ the $1d$ spectrum is singular-continuous, all $1d$
eigenfunctions $\psi_i(n)$ are critical, decaying as a power laws from
different points $n_0$. Any product of two such functions, as in
Eq.~(\ref{eq:product}), must be critical on the $2d$ quasicrystal as
well, yet we have observed that for $t$ close to 1 the spectrum is
continuous, implying that the wave functions must be extended.

\begin{figure}[tbh]
\hspace*{0pt}
\begin{center}

  \begin{tabular}{c c c}
  {\tiny (a) $1d$, $t=1.2$} & {\tiny (b) $1d$, $t=2.0$} 
    & {\tiny (c) $1d$, $t=3.0$}\\
  \scalebox{0.50}{\rotatebox{00}{\includegraphics*{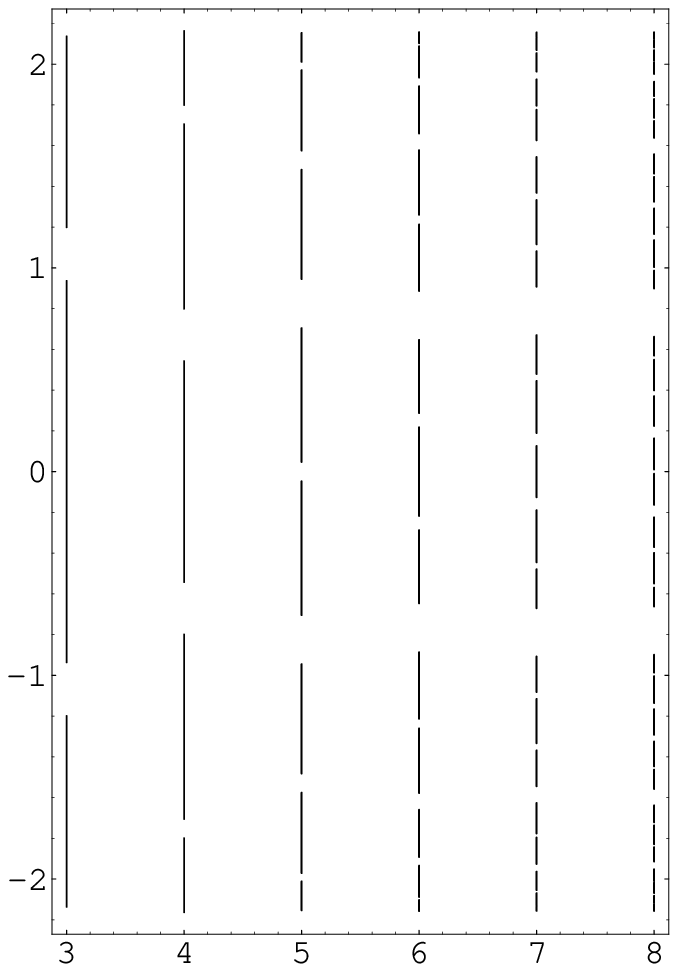}}} &
  \scalebox{0.50}{\rotatebox{00}{\includegraphics*{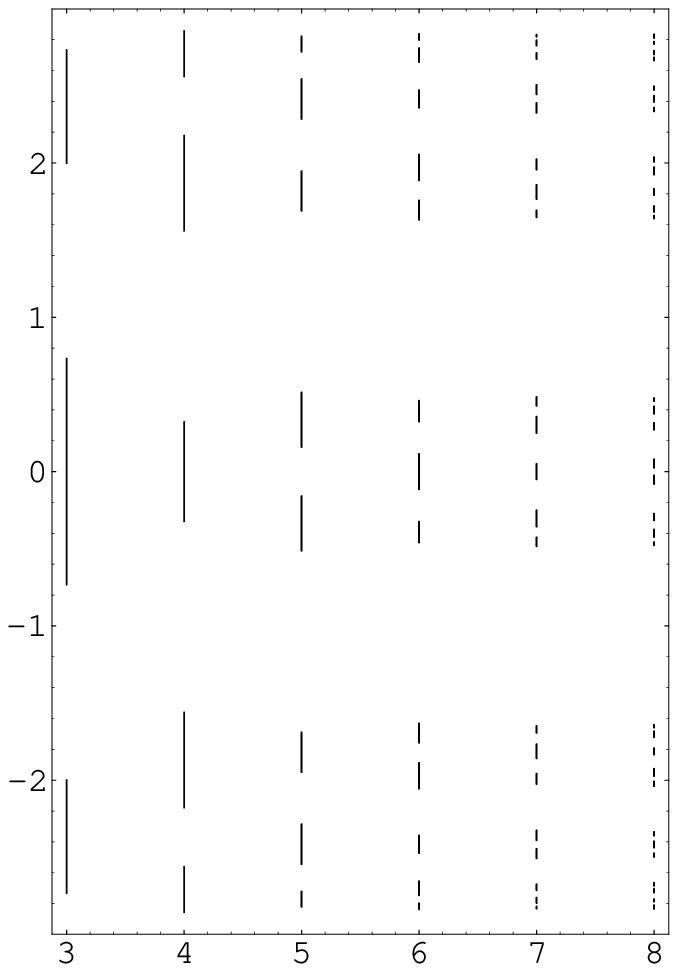}}} &
  \scalebox{0.50}{\rotatebox{00}{\includegraphics*{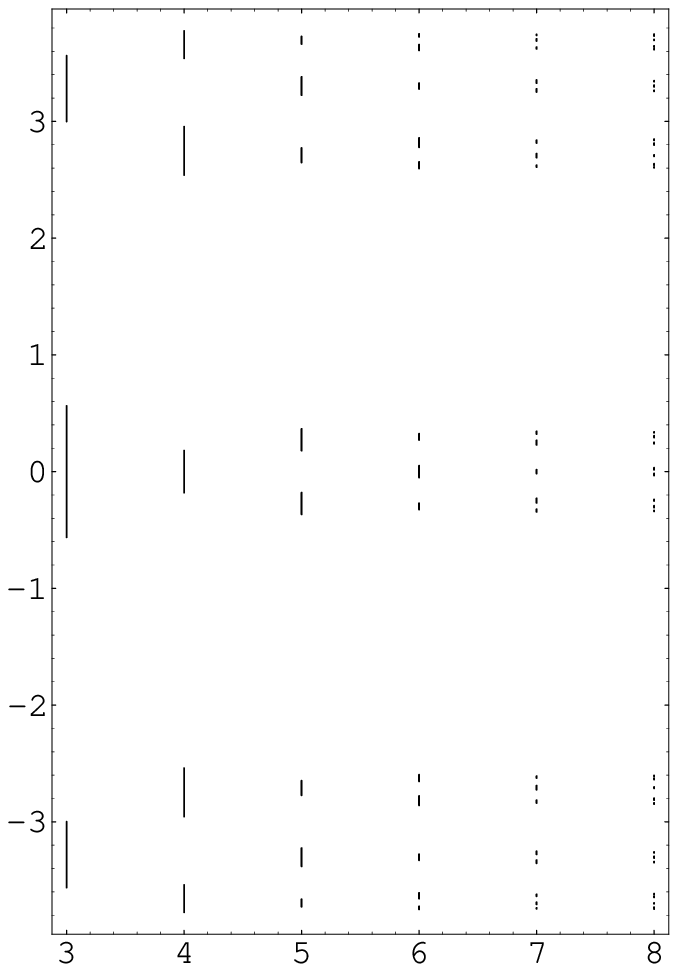}}} \\
  {\tiny (d) $2d$, $t=1.2$} & {\tiny (e) $2d$, $t=2.0$} 
    & {\tiny (f) $2d$, $t=3.0$}\\
  \scalebox{0.50}{\rotatebox{00}{\includegraphics*{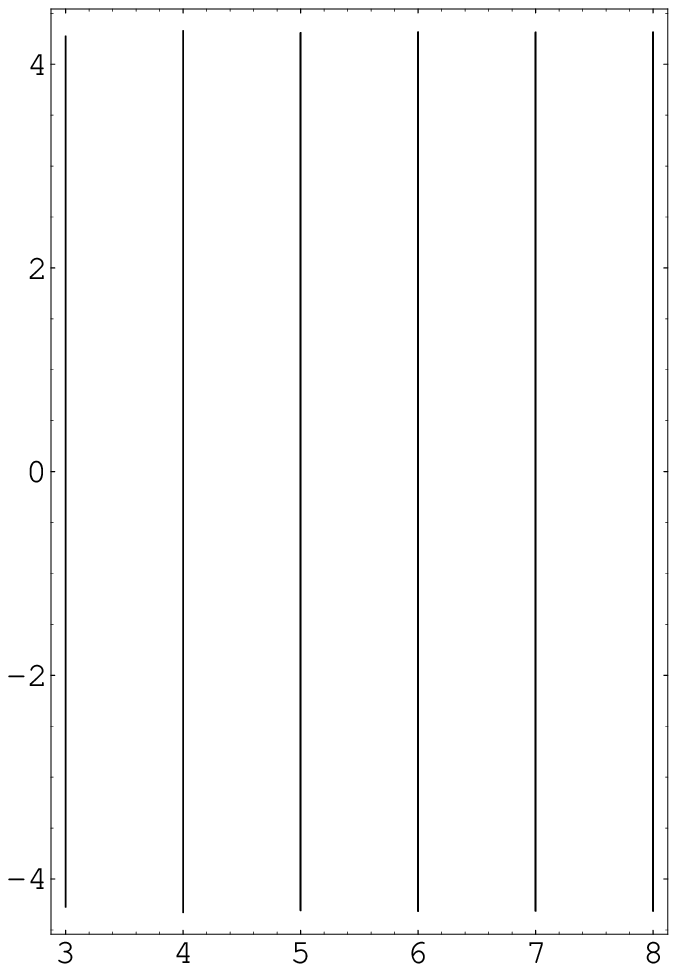}}} &
  \scalebox{0.50}{\rotatebox{00}{\includegraphics*{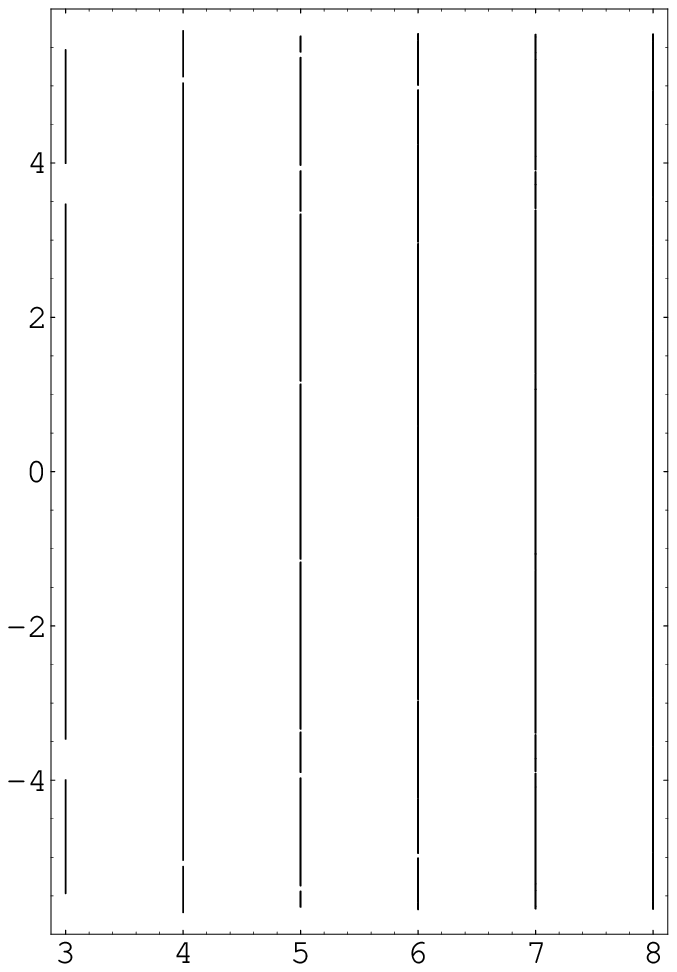}}} &
  \scalebox{0.50}{\rotatebox{00}{\includegraphics*{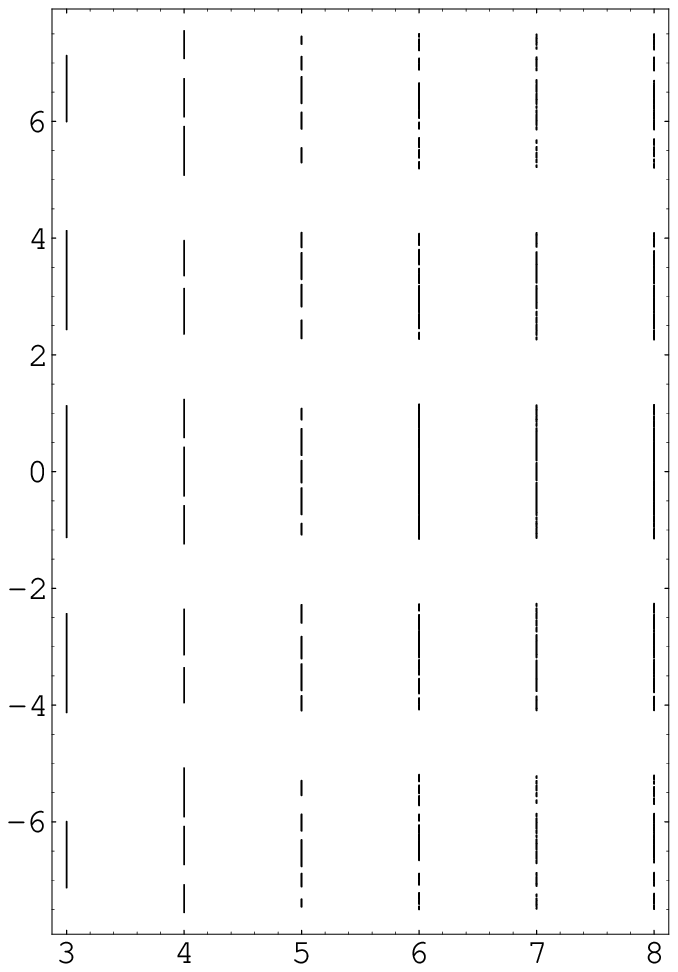}}} \\
  \end{tabular}


\parbox{10.cm}{\caption{Energy spectra of the first few $1d$ and $2d$
    approximants with $t=1.2$, $2.0$, and $3.0$. In (a)--(c) the
    number of bands for the 3$^{rd}$ to the 8$^{th}$ approximants are
    $3,5,8,13,21$, and 34. In (d) there is a single band for all
    approximants; in (e) the number of bands are $3,3,9,13,19$, and 31;
    and in (f) there are $5,13,23,47,87$, and 213 bands.}}

\end{center}
\end{figure} 

We would like to conjecture that the resolution of this paradox stems
from the high degeneracy of each eigenvalue when bands overlap and
merge into one as in Fig.~2(d). In general, many different pairs of
$1d$ eigenvalues $E_i$ and $E_j$ may add up to the same $2d$
eigenvalue $E$. Each eigenfunction $\Psi_{ij}(n,m)$ with eigenvalue
$E$ is critical, all peaked at different points $(n_0,m_0)$ on the
$2d$ quasicrystal, with substantial overlap due to their slow spatial
decay. It is plausible that one could construct linear combinations of
these critical eigenfunctions that are extended over the whole
infinite quasicrystal. We intend to investigate this conjecture in the
near future.

To support our conjecture, we conclude by showing that for any value
of $t$ the eigenfunctions with energy $E=0$ are extended over the $2d$
quasicrystal. First recall that for any $1d$ eigenfunction $\psi_i(n)$
with energy $E_i$, the function $(-1)^m\psi_i(m)$ is a $1d$
eigenfunction with energy $-E_i$.  Therefore, $\Psi_i(n,m) =
(-1)^m\psi_i(n)\psi_i(m)$ is a $2d$ eigenfunction with energy $E=0$.
Thus, for the $n^{th}$ approximant, the energy $E=0$ is $F_n$-fold
degenerate (a similar situation arises for the labyrinth
tiling~\cite{labyrinth}). Since the $1d$ eigenfunctions form a
complete set that spans all functions on the Fibonacci chain, one can
perform a change of basis to an alternative complete set
$\phi_j(n)=\sum_i c_{ji}\psi_i(n)$ whose members are all extended.
The $2d$ eigenfunctions $\Phi_j(n,m) = (-1)^m\phi_j(n)\phi_j(m)$, all
with energy $E=0$ are extended over the whole $2d$ quasicrystal.

We should conclude with a cautionary remark that all our conclusions rely
on the fact that the behavior for finite $2d$ approximants survives in
the infinite limit. This is not obvious until proven rigorously as in $1d$.

\rule{0.cm}{0.3cm}

\noindent {\bf Acknowledgment.} We thank Michael Baake for valuable
comments. This research is supported by the Israel Science Foundation
through Grant No.~278/00.



\begin{thebibliography}{00}
  
\bibitem{kohmoto} M.~Kohmoto, L.P.~Kadanoff, and C.~Tang, Phys. Rev.
  Lett. {\bf 50}, 1870 (1983); S.~Ostlund, R.~Pandit, D.~Rand,
  H.S.~Schellnhuber, and E.D.~Siggia, Phys. Rev. Lett. {\bf 50}, 1873
  (1983); M.~Kohmoto and J.R.~Banavar, Phys. Rev. B {\bf 34}, 563
  (1986); and M.~Kohmoto, B.~Sutherland, and C.~Tang, Phys. Rev. B {\bf
    35}, 1020 (1987).

\bibitem{reviews} T.~Janssen, in {\it The Mathematics of Long-Range
    Aperiodic Order}, ed. R.V.~Moody, Kluwer, Dordrecht, 269 (1997);
    T.~Fujiwara, in {\it Physical Properties of Quasicrystals},
    ed. Z.M.~Stadnik, Springer, Berlin, ch. 6 (1999); J.~Hafner
    and M.~Kraj{c}\'{\i}, {\it ibid.} ch. 7; and D.~Damanik, in {\it
    Directions in Mathematical Quasicrystals}, ed. M.~Baake and
    R.V.~Moody, AMS, Providence, 277 (2000).

\bibitem{sire} C.~Sire, in {\it Proc. ICQ5}, ed. C.~Janot and
  R.~Mosseri, World Scientific, Singapore, 415 (1995).

\bibitem{squarefib} R.~Lifshitz. J. of Alloys and Compounds {\bf 342},
  186 (2002); and R.~Lifshitz, Foundations of Physics, {\it in
    press,} (2003).

\bibitem{ueda} K.~Ueda and H.~Tsunetsugu, Phys. Rev. Lett. {\bf 58},
  1272 (1987); W.A.~Schwalm and M.K.~Schwalm, Phys. Rev. B {\bf 37},
  9524 (1988); J.A.~Ashraff, J-M.~Luck, and R.B.~Stinchcombe,
  Phys. Rev. B {\bf 41}, 4314 (1990); J.X.~Zhong and R.~Mosseri,
  J. Phys: Condens. Matter {\bf 7}, 8383 (1995); and S.~Roche and
  D.~Mayou, Phys. Rev. Lett. {\bf 79}, 2518 (1997).
  
\bibitem{labyrinth} C.~Sire, Europhys. Lett. {\bf 10}, 483 (1989); and
  H.Q.~Yuan, U.~Grimm, P.~Repetowicz, and M.~Schreiber, Phys. Rev. B
  {\bf 62}, 15569 (2000).

\end{thebibliography}
\end{document}